\documentclass[aps,reprint,superscriptaddress]{revtex4-1}
\usepackage[utf8]{inputenc}
\usepackage{graphicx}
\usepackage[dvipsnames]{xcolor}
\usepackage[sort&compress]{natbib}
\usepackage{booktabs}
\usepackage{bm}
\usepackage{amsmath}
\begin{document}

\title{From spherical chicken to the tipping points of a complex network} 

\author{Gui-Yuan Shi} \email{guiyuan.shi@unifr.ch}
\affiliation{Niels Bohr Institute, University of Copenhagen, Copenhagen, Denmark}
\affiliation{Department of Physics, University of Fribourg, Fribourg, Switzerland}
\author{Rui-Jie Wu}
\affiliation{Niels Bohr Institute, University of Copenhagen, Copenhagen, Denmark}
\author{Yi-Xiu Kong}
\affiliation{Department of Physics, University of Fribourg, Fribourg, Switzerland}
\author{Kim Sneppen}
\affiliation{Niels Bohr Institute, University of Copenhagen, Copenhagen, Denmark}

\date{\today}

\begin{abstract}
The outbreak of epidemics, the emergence of the financial crisis, the collapse of ecosystem, and the explosive spreading of rumors,
we face many challenges in today's world. 
These real-world problems can be abstracted into a sequential break down of a complex system in an increasingly stressful environment. Because both the system and the environment require a large number of parameters to describe, 
the break down conditions has been difficult to estimate. 
We use a highly symmetric system to gauge a complex environment, which  enables us to propose a scalar benchmark to describe the environment. 
This allows us to prove that all the tipping points of a complex network fall between the maximum k-core and maximum eigenvalue of the network.
\end{abstract}
\maketitle

We start with the story about a physicist and a spherical chicken. When winter comes, a farmer's chickens all get sick and the farmer does not know what is wrong with them. The farmer calls his neighbour, a physicist, to see if he can figure out what is wrong. The physicist looks at the chickens and then starts scribbling in a notebook. Finally, after several gruesome calculations, he exclaims, "I have got it! But it only works for spherical chickens."

Let us consider his model seriously. The model assumes a chicken would feel too cold because of the combined effects of air temperature($t_a$), relative humidity($h_r$), wind speed($v_w$) and sunlight intensity($I_s$). These spherical chickens are put into an isotropic environment, and the radius of the spherical chickens is the only parameter to describe them. It is not hard to imagine that the larger the radius, the more cold-resistant the chickens are.

\begin{figure}[htb]
\centering
\includegraphics[width=8cm]{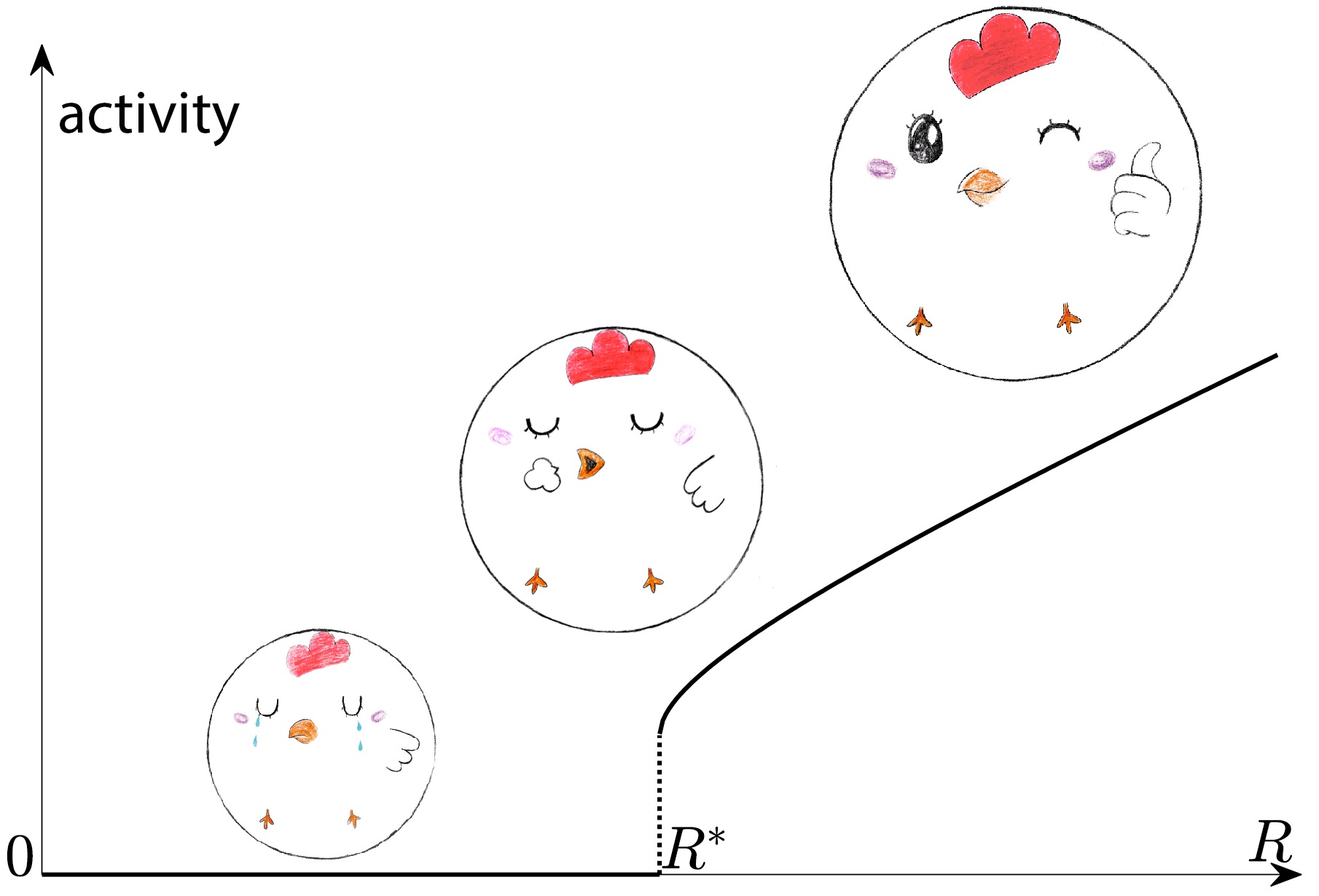}
\caption{
Given an environment, the activity of a spherical chicken increases with its radius.
}
\label{chicken_fig}
\end{figure}

Therefore, for a given environment, we find the smallest spherical chicken that can endure the coldness, its radius can be expressed as a function of the environment parameters $(t_a,h_r,v_w,I_s)$. We use the critical radius $R^*(t_a,h_r,v_w,I_s)$ to represent the effective coldness of the given environment. If we draw contour lines for $R^*$ in the 4-D parameter space $(t_a,h_r,v_w,I_s)$, one can imagine that a real chicken will feel differently along a contour line. Because the real chicken might be more windproof than a spherical chicken, but worse at absorbing sunlight.

Thus, for a given real chicken, we want to figure out the environmental regime that makes the chicken feel not cold. For the boundary of the regime, there exists a lowest effective coldness $R_l$ and the highest $R_h$ among the contour lines for $R^*(t_a,h_r,v_w,I_s)$.
That is to say, if we put the chicken in an environment that effective coldness $R^*$ 
that is higher than $R_h$, for sure it will feel too cold; and if $R^* < R_l$, 
the chicken will never feel cold. 
In practice the upper and lower bounds of effective coldness endurance, $R_h$ and  $R_l$ for any given chicken are important to know.

\begin{figure*}[htb]
\centering
  \begin{tabular}{@{}ccc@{}}
    \includegraphics[width=12cm]{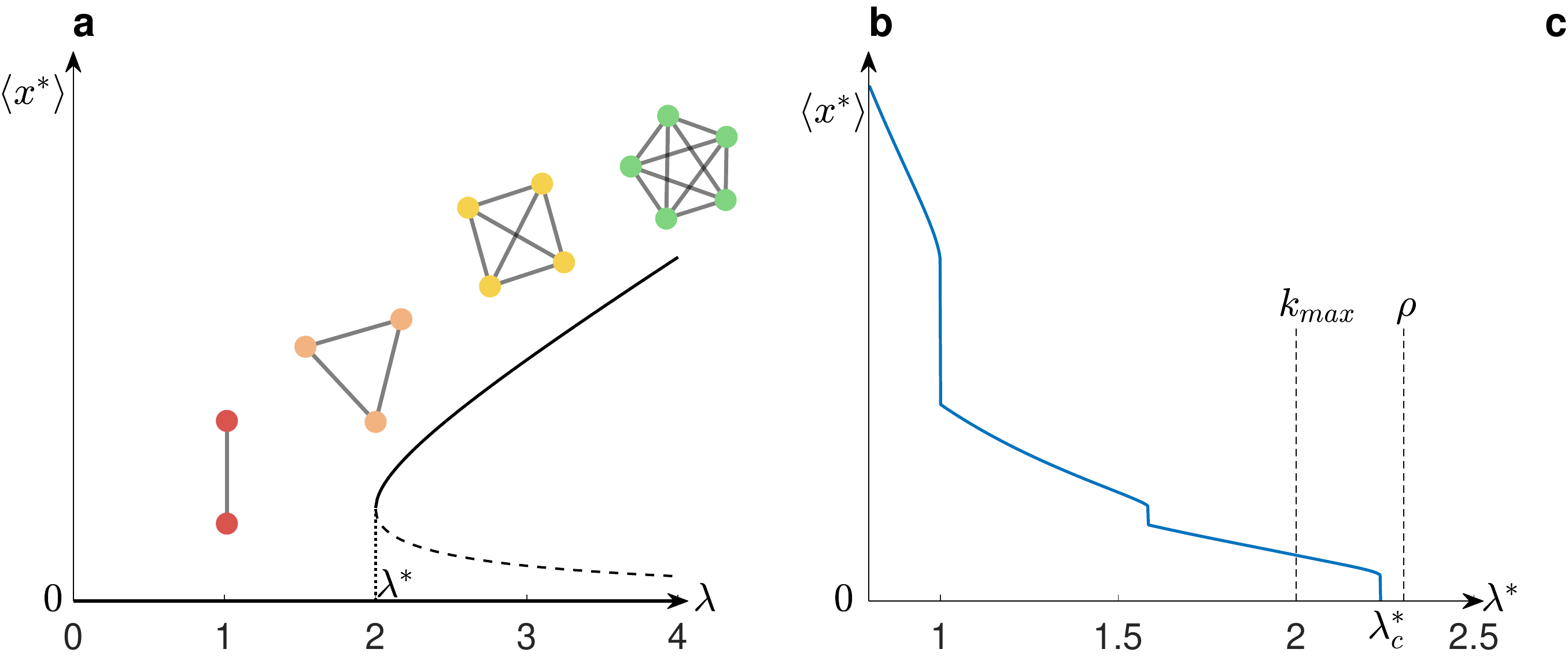}&
    \includegraphics[width=5cm]{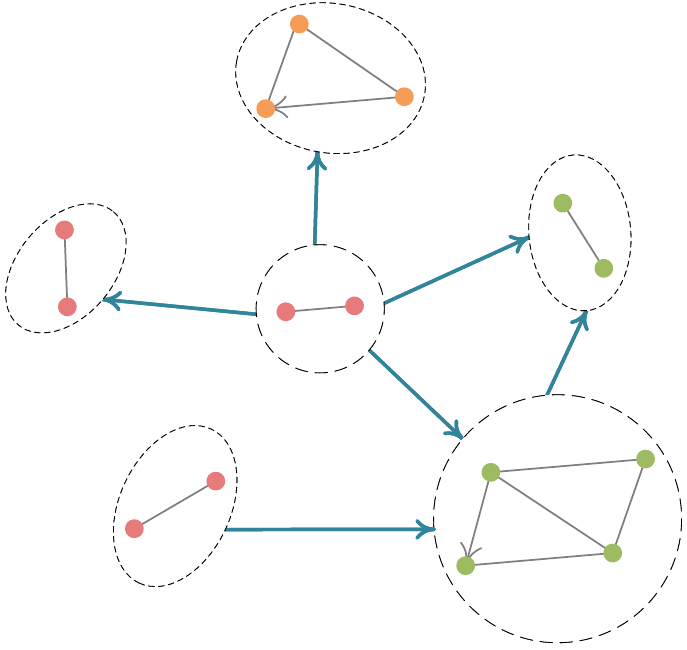}
      \end{tabular}
  \caption{
An illustration of $\lambda^*$ and $\lambda_c^*$. The vertical axis $\langle x^* \rangle $shows the average abundance of components, indicating whether or not the system collapses. 
\textbf{a}) For homogeneous networks, $\lambda = \sum_j{A_{ij}}$. Given the functional form $F$ and set of environmental parameters $\alpha$, $\lambda$ increases as the density of the network increases and the system is more likely to have a non-zero fixed-point solution. For a specific $F$ and $\alpha$, the minimum $\lambda$ that allows the existence of the non-zero solution is marked as $\lambda^*$. 
\textbf{b}) For {\it E.coli} network $A$ , if we fix the network and tune the environmental parameters so that it is increasingly difficult for the system to survive(increasing $\lambda^*$), and our theory predicts all the tipping points will emerge between $\lambda^*_{min} = k_{max}(A)$ and $\lambda^*_{max} = \rho(A)$.
\textbf{c}) We condense the {\it E.coli} network by contracting $6$ strongly connected subgraph of the original graph into $6$ giant nodes. During the evolution the red nodes go extinct at $\lambda^*=1$, orange nodes go extinct at $\lambda^*=1.58$, and green nodes go extinct at $\lambda^*=2.24$.
Because the nodes support each other within a strongly connected subgraph, and can also contribute to the downstream nodes but not to upstream nodes.
}
\label{lambda_fig}
\end{figure*}

Now we go back to the study of complex network, described by following set of differential equations
\begin{equation}
\frac{dx_i}{dt}=F(x_i, \sum_{j=1}^{N}A_{ij}G(x_i,x_j)),
\label{general_eq}
\end{equation}
here $x_i$ is the abundance of component $i$, the weighted adjacency matrix $A_{ij}\geq0$ captures the support strength from $j$ to $i$.
The first term of $F$ indicates the self-decay of $i$ in isolation, and the second term describes the total support from its direct neighbors in the network.
That is, following previous works \cite{gao2016universal,morone2019k} here we also focus on cooperative systems where each part of the system is contributing positively to existence of other parts (both $F$ and $G$ is increasing as function of $x_j$, see method for a detailed analysis for properties of the equations we considered).

With an appropriate choice of $F$ and $G$, Eq.~\ref{general_eq} can be used to describe numerous systems include mutualistic ecosystem, disease spreading, information propagation, collaboration network and so on.
We can imagine that a difficult environment would cause the decay to be too fast, or the support to be inefficient, then the whole system can only stay in zero abundance state. However, increasing the connectivity or coupling strength in the complex network $A$ may allow the system to work (and our chicken to lay eggs).

Scientists have made efforts in finding the tipping points from many approaches, 
such as using mean-field approximation\cite{pastor2001epidemic, barrat2008dynamical},
linear stability analysis\cite{lajmanovich1976deterministic},
second-order mean-field under quasi linear assumption \cite{gao2016universal},
and logistic approximation \cite{morone2019k}. 

Here we will use a framework inspired by our spherical chickens. 
First we use a homogeneous network with constant in-degree $\sum_{j}A_{ij} = \lambda$ (a spherical chicken with radius $R$) to gauge the effective difficulty level $\lambda^*$ (as effective coldness $R^*$) of an environment. 
This is done through the 1D equation (simply replace $x_i$, $x_j$ with $x$ and replace $\sum_{j}A_{ij}$ with $\lambda$)
\begin{equation}
F(x, \lambda G(x,x)) = 0,
\label{eff_eq}
\end{equation}
obtained from Eq.~\ref{general_eq} when all nodes have symmetrical input. The effective difficulty level $\lambda^*$ is the minimal $\lambda$ that enable $x$ to have positive solution (see Fig.~\ref{lambda_fig}\textbf{a}).

Given a network $A$, each of its tipping points has a corresponding $\lambda^*$, denoted as $\lambda^*_c$. We prove (see methods) that for any given network $A$, the lower and upper bounds of $\lambda^*_c$ (see Fig.~\ref{lambda_fig}\textbf{b}) have to be limited by properties of the network $A$:
\begin{equation}
   k_{max}(A) \leq \lambda^*_c \leq \rho(A).
\end{equation}
Here $k_{max}(A)$ is an extension of the traditional maximum $k$-core to the weighted networks~\cite{eidsaa2013s, kong2019k}.
It is defined as the maximum number that allows the existence of a subgraph, in which each node receive at least $k_{max}(A)$ weighted incoming edges in total within the subgraph;
and $\rho(A)$ is the spectral radius, i.e. the largest eigenvalue, of $A$.
Note that the largest eigenvalue $\rho(A)$ is always larger than $k_{max}(A)$~\cite{shin2016corescope}. 

Further, if the interaction term $G(x_i,x_j)$ is a step function of $x_j$, the tipping points always follows $\lambda^* = k_{max}(A)$. In contrast, when Eq.~\ref{eff_eq}'s solution from zero to nonzero undergoes a continuous transition, all the transition points satisfy $\lambda^* = \rho(A)$.

In Fig.~\ref{lambda_fig}\textbf{c} we show the abstracted graph of the {\it E.coli} network. With this illustration we can explain the "drops" at  $\lambda^*=1$ and $\lambda^*=1.58$ as shown in Fig.~\ref{lambda_fig}\textbf{b}. These two drops are also "tipping points" and can be studied by our theory. But for simplicity, we only analyze the tipping points that result in a total collapse of the network. In the following, we give a few examples to illustrate our theory.

\textbf{Gene regulatory networks.}
\begin{figure*}[htb]
\centering
\includegraphics[width=17cm]{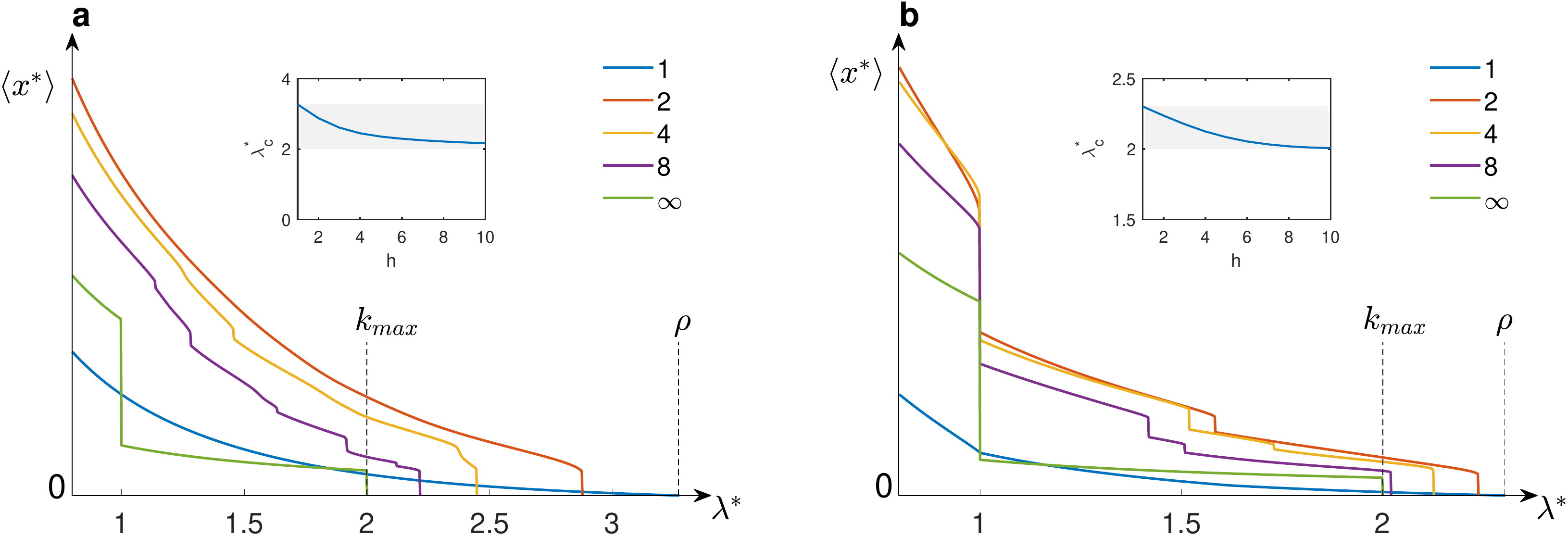}
\caption{
\textbf{The tipping points of two gene regulatory networks} Here $h = 1, 2, 4, 8, \infty$. The two dashed lines represent $k_{max}$(left) and $\rho$(right) respectively.  The inset shows the changes of $\lambda_c^*$ depending on the parameter $h$, and the grey area shows the predicted bounds of $\lambda_c^*$.
We recursively pruned the nodes whose in-degree or out-degree equals zero to run the simulation faster, because these nodes do not affect the tipping points. We use $\beta_{\mathrm{eff}}^{origin}$ to denote the $\beta_{\mathrm{eff}}$ of the network before pruning.
\textbf{a}) {\it S. cerevisiae}~\cite{balaji2006comprehensive}.  
$k_{max} = 2$, $\rho = 3,27$, $\beta_{\mathrm{eff}}^{origin} = 3,43$ (4441 nodes), $\beta_{\mathrm{eff}} = 3.30$ (60 nodes).
\textbf{b}) {\it E. coli}~\cite{gama2008regulondb}.
$k_{max} = 2$, $\rho = 2.30$, $\beta_{\mathrm{eff}}^{origin} = 1.17$ (1550 nodes), $\beta_{\mathrm{eff}} = 2.21$ (15 nodes). 
Data from https://github.com/jianxigao/NuRsE}
\label{gene_fig}
\end{figure*}
First let us consider the example of 
gene regulatory networks. As in the analysis of Ref~\cite{gao2016universal},
we wrongly assume that all inputs are positive, i.e. that all are activation regulations of type:
\begin{equation}
\frac{dx_i}{dt}=-Bx_i+\sum_{j=1}^{N}A_{ij}\frac{x_j^h}{1+x_j^h}.
\label{gene_eq}
\end{equation}
Here the first term determines the degradation, whereas the parameter $h$ is the Hill coefficient that quantifies cooperativity of the gene regulation. 
The corresponding 1D equation of Eq.~\ref{gene_eq} is: 
\begin{equation}
-B x +\lambda \frac{x^h}{1+x^h}=0.
\end{equation}
Obviously when $h<1$ a positive solution always exists. When $h \geq 1$, the effective difficulty level $\lambda^*$ is (note $0^0 = 1$)
\begin{equation}
\lambda^* = Bh(h-1)^{\frac{1}{h}-1}.
\end{equation} 
$h$ is fixed by the details of the biological 
regulation, whereas $B$ changes dependent on external stress/living conditions. 
For each $B$, we can calculate the effective difficulty $\lambda^*$, and have the corresponding average abundance $\langle x^* \rangle$ through numerical simulation.(Fig.~\ref{gene_fig})
At a tipping point $\lambda^* = \lambda^*_c$, the corresponding $\langle x \rangle$ collapse to zero.

We predict that $\lambda_c^*$ will change from $\rho(A)$ (when $h = 1$) to $k_{max}(A)$ (when $h\to+\infty$).
In comparison, Gao et~al. suggest that $\lambda_c^* = {\beta_{\mathrm{eff}}\equiv \langle s^{out}s^{in} \rangle / \langle s \rangle}$, independent of $h$. Here $s$ is degree and $\langle s\rangle = \langle s^{out}\rangle= \langle s^{in}\rangle$. 
In many situations, $\beta_{\mathrm{eff}} \approx \rho(A)$, see Ref.~\cite{chung2003spectra,castellano2017relating}. 
But in certain cases, they are quite different. 
Imagine two extreme case:
(i) if we add a new node whose in-degree are very high but out-degree equals zero, then $\beta_{\mathrm{eff}}$ will decrease significantly while the largest eigenvalue and the tipping points will not change. Small $\beta_{\mathrm{eff}}$ indicates a overestimation of collapse risk. {\it E. coli} network is an example(Fig.~\ref{gene_fig}\textbf{b});
(ii)for star network, $\beta_{\mathrm{eff}} = N/2$, $\rho = \sqrt{N-1}$. In this case $\beta_{\mathrm{eff}}$ underestimates the risk. Later we will see the latter case can explain phenomena observed in mutualistic networks.

Our simulations on two networks {\it S. cerevisiae} and {\it E.coli} are shown in Fig.~\ref{gene_fig}. 
The figure illustrates tipping points for different $h$ values, 
with $G$ varying from first order to a step function.
It is easily seen that all tipping points fall within our predicted region, and that the tipping point changes with onset of non-linearity, thus differing from the quasi linear estimates of Ref.\cite{gao2016universal}.

In Fig.~\ref{gene_fig}, both networks collapse at $\lambda^* = 1$ when $h\to\infty$. At this limit nodes with abundance $x^*<1$ does not contribute to their downstream targets.
Besides, in Fig.~\ref{gene_fig}\textbf{b}, the collapses at $1$ and near $1.5$ reflect the network has several strongly connected graphs with different tipping points (Fig.~\ref{lambda_fig}\textbf{c}).


\textbf{Mutualistic ecosystems.}
\begin{figure*}[htb]
\centering
\includegraphics[width=17cm]{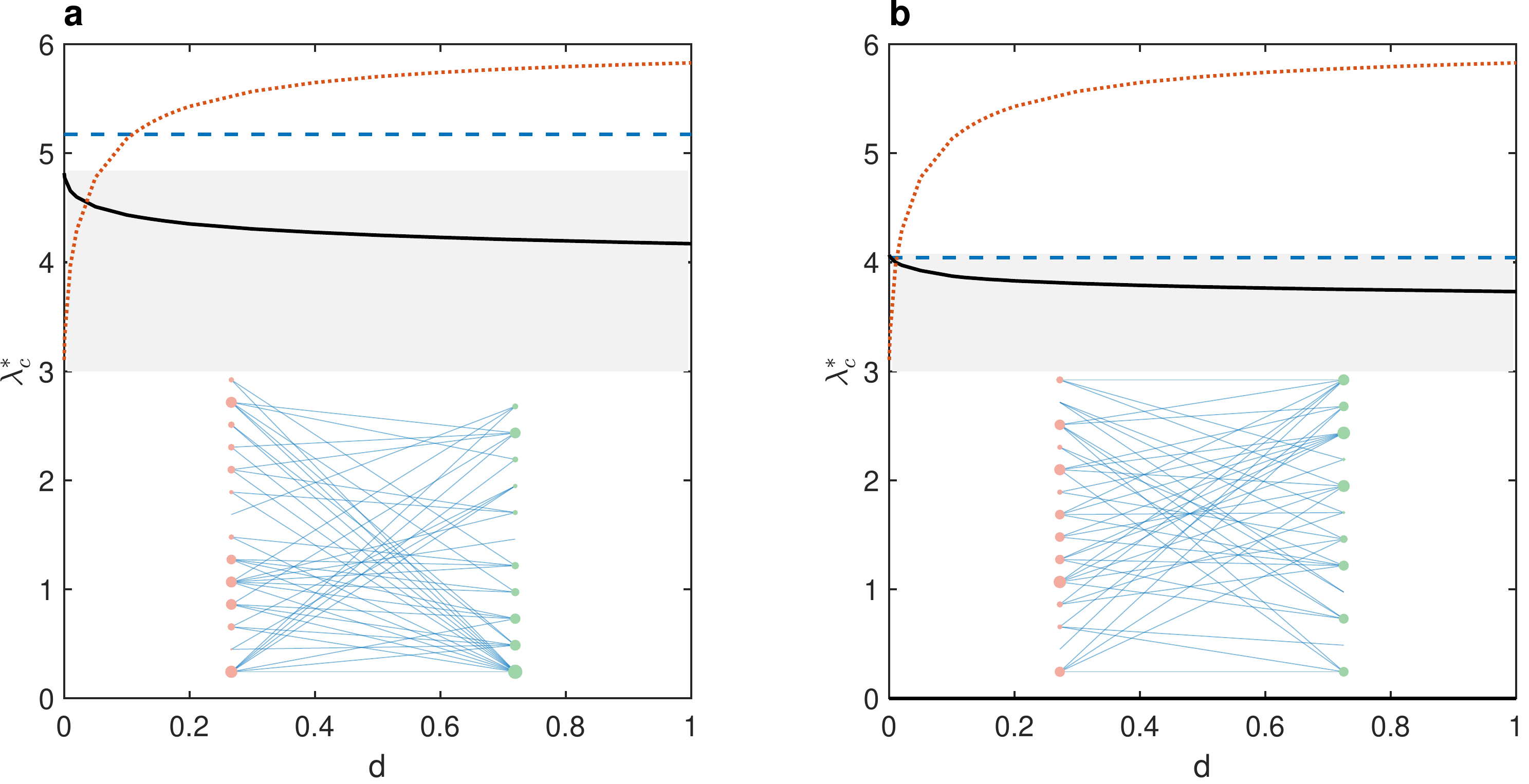}
\caption{
\textbf{Critical $\lambda$ of a real mutualistic network and a random network.}
We fix $s = 1$, $a = 1$. For each $d$, we find the critical $\gamma$, and then calculate $\lambda_c^*$ by Eq.~\ref{mutual_lambda}.
The insets show the structure of the two networks. Red nodes represent the birds and green nodes are the plants. The size of a node shows the abundance of the corresponding species near tipping point. 
The solid black line shows our simulation result of $\lambda_c^*$ relative to the parameter $d$, the grey area indicates the predicted bounds of $\lambda_c^*$, 
the dashed blue line shows the prediction $\beta_{\mathrm{eff}}$ in Ref.~\cite{gao2016universal}, 
and the dotted red line shows the prediction in Ref.~\cite{morone2019k}. \textbf{a})A real mutualistic network. The network contains $14$ birds, $12$ plants, $46$ links~\cite{sorensen1981interactions}, data from https://www.nceas.ucsb.edu/interactionweb/.
\textbf{b}) The random network generated by randomly swapping the links of network in \textbf{a}. We can find that $\beta_{\mathrm{eff}}$ roughly equals to the upper bound $\rho(A)$, which suggests Gao et al's~\cite{gao2016universal} results are valid for random networks when transition is continuous ($\lambda_c^*=\rho(A)$).}
\label{mutual_fig}
\end{figure*}
Morone et~al.~\cite{morone2019k} studied a mutualistic ecosystem governed by: 
\begin{equation}
\frac{dx_i}{dt}=-x_id-x_i^2s+\frac{ \gamma \sum_{j=1}^{N} A_{ij} x_j }{ \alpha + \sum_{j=1}^{N}A_{ij}x_j}x_i
\end{equation}
Here $d>0$ is the death rate, $s>0$ is a self-limitation parameter, $\alpha>0$ is the half-saturation constant, and $\gamma>0$ is the mutualistic interaction efficiency.
They adopted a logistic approximation that assume $x/(\alpha + x)$ equals 0 when $x<\alpha$ and 1 when $x>\alpha$, 
proposed the threshold on the mutualistic benefit
$K_\gamma = \frac{\alpha s (\gamma+d)}{(\gamma-d)^2}$, 
and predicted that all tipping points satisfy $K_\gamma = k_{max}(A)$.

Here we study the 1D equation
\begin{equation}
-x d-x^2 s+\frac{\gamma \lambda x^2}{\alpha + \lambda x} = 0,
\end{equation}
which has a critical minimal $\lambda^*$ that allows for positive $x$ 
\begin{equation}
\lambda^* = \frac{\alpha s}{(\sqrt{\gamma}-\sqrt{d})^2}.
\label{mutual_lambda}
\end{equation}
This $\lambda^*$ is coincident with $K_\gamma$ 
when $d = 0$.
In this case, Morone et~al~\cite{morone2019k} predicted the tipping points satisfy $k_{max}(A) = \alpha s/\gamma$.
Our methodology further teaches us (see methods) that for $d=0$, the transition is continuous around $\rho(A) = \alpha s/\gamma$.

Using a mutualistic network we illustrate our result in Fig.~\ref{mutual_fig}\textbf{a}.
We fix $s =1$ and $a = 1$, and plot $\lambda_c^*$ as function of parameter $d$ from $0$ to $1$.
We see that $\lambda_c^*$ are between $k_{max}(A)$ and $\rho(A)$, just as we predicted. As $d$ increases, $\lambda^*_c$ will decrease starting from $\rho(A)$. 
In contrast Morone et~al.'s results lead to $\lambda^*_c$ increase starting from $k_{max}(A)$ to $2k_{max}(A)$. 
This means for small $d$, they overestimate the risk of ecosystem collapse; and when $d$ is large, they underestimate the risk. 
The systematic biases are also observed in their simulation (Fig.~2g in Ref.~\cite{morone2019k}).
Also, $\beta_{\mathrm{eff}}$ from Ref.~\cite{gao2016universal} often lies beyond the upper bound of the critical area, indicate an underestimation of the collapse risk. 
This is due to the existence of a large hub that distorts their result(recall the star network we discussed above).

In addition, the largest eigenvalue $\rho (A)$ itself is also a measure of nestedness~\cite{staniczenko2013ghost,mariani2019nestedness}, i.e. a highly nested network tends to have a large $\rho (A)$.
As $\lambda_c^*$ is highly correlated with $\rho (A)$, the nested networks will have larger $\lambda_c^*$ and survive easier at higher stress.
This is consistent with the observation~\cite{lever2014sudden, rohr2014structural, saavedra2016nested} that highly nested networks are more robust. 

Further by comparing Fig.~\ref{mutual_fig} \textbf{a} and \textbf{b}, we can see that the real network has a larger $\lambda_c^*$ than random network, suggesting that the real networks are more robust than random networks.

\textbf{Epidemic process.}
Finally we study the SIS model in epidemic process~\cite{pastor2015epidemic},
\begin{equation}
\frac{dx_i}{dt} = -x_i  + \beta (1-x_i) \sum_{j=1}^{N} A_{ij} x_j
\end{equation}
Here $\beta>0$ is the effective transmission rate. The corresponding 1D equation is
$-x + \lambda \beta (1-x) x = 0$,
and the effective difficulty level
$\lambda^* = 1/\beta$ (for sustained endemic).
We predict the transition is continuous near $\beta\cdot\rho(A) = 1$(see methods), which is in agreement with the former studies~\cite{pastor2015epidemic}. 

\textbf{Discussion.}
In conclusion, we have proposed a universal framework to study the tipping points of a complex system. First, we use homogeneous networks (spherical chickens) as a benchmark to gauge the difficulty level of an environment (determined by the parameter sets). Due to the symmetry of the nodes in the homogeneous network, for any given environment we can just study the 1D equation to find the sparsest homogeneous network that can survive in the environment. The in-degree of this sparsest homogeneous network $\lambda^*$ can be regarded as the effective difficulty level of the environment, and can usually be studied analytically. 

Normally a heterogeneous network will have different resilience under different environmental parameter set corresponding to an identical effective difficulty level. We prove that for any given network $A$, the effective difficulty level of all the tipping points $\lambda_c^*$ are between its maximum $k$-core $k_{max}(A)$ and its largest eigenvalue $\rho(A)$. That is to say this system can always survive in effective difficulty level $\lambda^* = k_{max}(A)$, and will always collapse in $\lambda^* = \rho(A)$. This is particularly meaningful when in reality we need to make the network functions or malfunctions under certain conditions. With the knowledge of the upper bounds and lower bounds, we will be able to control the parameters/network that can keep the system in a desirable state. In some special case(step function or continuous transition), all the tipping points of a heterogeneous system have an identical effective difficulty level $k_{max}(A)$ or $\rho(A)$. In these case, we can even know all the explicit tipping points. 

In short, our new findings theoretically determine the bounds of the tipping points of a large class of dynamical systems, and our results unveil important information for controlling the networks in real world.
The methodology we developed is general and may be used in other scientific disciplines, especially when the problem can be abstracted as a complex object(real chicken)/group(heterogeneous network) evolving under an environment described by multiple parameters.

\textbf{Acknowledgements}
This project has received funding from the European Research Council (ERC) under the European Union's Horizon 2020 research and innovation program under grant agreement No 740704.

\bibliographystyle{apsrev4-1}
\bibliography{bibliography}

\begin{thebibliography}{22}%
\makeatletter
\providecommand \@ifxundefined [1]{%
 \@ifx{#1\undefined}
}%
\providecommand \@ifnum [1]{%
 \ifnum #1\expandafter \@firstoftwo
 \else \expandafter \@secondoftwo
 \fi
}%
\providecommand \@ifx [1]{%
 \ifx #1\expandafter \@firstoftwo
 \else \expandafter \@secondoftwo
 \fi
}%
\providecommand \natexlab [1]{#1}%
\providecommand \enquote  [1]{``#1''}%
\providecommand \bibnamefont  [1]{#1}%
\providecommand \bibfnamefont [1]{#1}%
\providecommand \citenamefont [1]{#1}%
\providecommand \href@noop [0]{\@secondoftwo}%
\providecommand \href [0]{\begingroup \@sanitize@url \@href}%
\providecommand \@href[1]{\@@startlink{#1}\@@href}%
\providecommand \@@href[1]{\endgroup#1\@@endlink}%
\providecommand \@sanitize@url [0]{\catcode `\\12\catcode `\$12\catcode
  `\&12\catcode `\#12\catcode `\^12\catcode `\_12\catcode `\%12\relax}%
\providecommand \@@startlink[1]{}%
\providecommand \@@endlink[0]{}%
\providecommand \url  [0]{\begingroup\@sanitize@url \@url }%
\providecommand \@url [1]{\endgroup\@href {#1}{\urlprefix }}%
\providecommand \urlprefix  [0]{URL }%
\providecommand \Eprint [0]{\href }%
\providecommand \doibase [0]{http://dx.doi.org/}%
\providecommand \selectlanguage [0]{\@gobble}%
\providecommand \bibinfo  [0]{\@secondoftwo}%
\providecommand \bibfield  [0]{\@secondoftwo}%
\providecommand \translation [1]{[#1]}%
\providecommand \BibitemOpen [0]{}%
\providecommand \bibitemStop [0]{}%
\providecommand \bibitemNoStop [0]{.\EOS\space}%
\providecommand \EOS [0]{\spacefactor3000\relax}%
\providecommand \BibitemShut  [1]{\csname bibitem#1\endcsname}%
\let\auto@bib@innerbib\@empty
\bibitem [{\citenamefont {Gao}\ \emph {et~al.}(2016)\citenamefont {Gao},
  \citenamefont {Barzel},\ and\ \citenamefont
  {Barab{\'a}si}}]{gao2016universal}%
  \BibitemOpen
  \bibfield  {author} {\bibinfo {author} {\bibfnamefont {J.}~\bibnamefont
  {Gao}}, \bibinfo {author} {\bibfnamefont {B.}~\bibnamefont {Barzel}}, \ and\
  \bibinfo {author} {\bibfnamefont {A.-L.}\ \bibnamefont {Barab{\'a}si}},\
  }\href@noop {} {\bibfield  {journal} {\bibinfo  {journal} {Nature}\ }\textbf
  {\bibinfo {volume} {530}},\ \bibinfo {pages} {307} (\bibinfo {year}
  {2016})}\BibitemShut {NoStop}%
\bibitem [{\citenamefont {Morone}\ \emph {et~al.}(2019)\citenamefont {Morone},
  \citenamefont {Del~Ferraro},\ and\ \citenamefont {Makse}}]{morone2019k}%
  \BibitemOpen
  \bibfield  {author} {\bibinfo {author} {\bibfnamefont {F.}~\bibnamefont
  {Morone}}, \bibinfo {author} {\bibfnamefont {G.}~\bibnamefont {Del~Ferraro}},
  \ and\ \bibinfo {author} {\bibfnamefont {H.~A.}\ \bibnamefont {Makse}},\
  }\href@noop {} {\bibfield  {journal} {\bibinfo  {journal} {Nature physics}\
  }\textbf {\bibinfo {volume} {15}},\ \bibinfo {pages} {95} (\bibinfo {year}
  {2019})}\BibitemShut {NoStop}%
\bibitem [{\citenamefont {Pastor-Satorras}\ and\ \citenamefont
  {Vespignani}(2001)}]{pastor2001epidemic}%
  \BibitemOpen
  \bibfield  {author} {\bibinfo {author} {\bibfnamefont {R.}~\bibnamefont
  {Pastor-Satorras}}\ and\ \bibinfo {author} {\bibfnamefont {A.}~\bibnamefont
  {Vespignani}},\ }\href@noop {} {\bibfield  {journal} {\bibinfo  {journal}
  {Physical review letters}\ }\textbf {\bibinfo {volume} {86}},\ \bibinfo
  {pages} {3200} (\bibinfo {year} {2001})}\BibitemShut {NoStop}%
\bibitem [{\citenamefont {Barrat}\ \emph {et~al.}(2008)\citenamefont {Barrat},
  \citenamefont {Barthelemy},\ and\ \citenamefont
  {Vespignani}}]{barrat2008dynamical}%
  \BibitemOpen
  \bibfield  {author} {\bibinfo {author} {\bibfnamefont {A.}~\bibnamefont
  {Barrat}}, \bibinfo {author} {\bibfnamefont {M.}~\bibnamefont {Barthelemy}},
  \ and\ \bibinfo {author} {\bibfnamefont {A.}~\bibnamefont {Vespignani}},\
  }\href@noop {} {\emph {\bibinfo {title} {Dynamical processes on complex
  networks}}}\ (\bibinfo  {publisher} {Cambridge university press},\ \bibinfo
  {year} {2008})\BibitemShut {NoStop}%
\bibitem [{\citenamefont {Lajmanovich}\ and\ \citenamefont
  {Yorke}(1976)}]{lajmanovich1976deterministic}%
  \BibitemOpen
  \bibfield  {author} {\bibinfo {author} {\bibfnamefont {A.}~\bibnamefont
  {Lajmanovich}}\ and\ \bibinfo {author} {\bibfnamefont {J.~A.}\ \bibnamefont
  {Yorke}},\ }\href@noop {} {\bibfield  {journal} {\bibinfo  {journal}
  {Mathematical Biosciences}\ }\textbf {\bibinfo {volume} {28}},\ \bibinfo
  {pages} {221} (\bibinfo {year} {1976})}\BibitemShut {NoStop}%
\bibitem [{\citenamefont {Eidsaa}\ and\ \citenamefont
  {Almaas}(2013)}]{eidsaa2013s}%
  \BibitemOpen
  \bibfield  {author} {\bibinfo {author} {\bibfnamefont {M.}~\bibnamefont
  {Eidsaa}}\ and\ \bibinfo {author} {\bibfnamefont {E.}~\bibnamefont
  {Almaas}},\ }\href@noop {} {\bibfield  {journal} {\bibinfo  {journal}
  {Physical Review E}\ }\textbf {\bibinfo {volume} {88}},\ \bibinfo {pages}
  {062819} (\bibinfo {year} {2013})}\BibitemShut {NoStop}%
\bibitem [{\citenamefont {Kong}\ \emph {et~al.}(2019)\citenamefont {Kong},
  \citenamefont {Shi}, \citenamefont {Wu},\ and\ \citenamefont
  {Zhang}}]{kong2019k}%
  \BibitemOpen
  \bibfield  {author} {\bibinfo {author} {\bibfnamefont {Y.-X.}\ \bibnamefont
  {Kong}}, \bibinfo {author} {\bibfnamefont {G.-Y.}\ \bibnamefont {Shi}},
  \bibinfo {author} {\bibfnamefont {R.-J.}\ \bibnamefont {Wu}}, \ and\ \bibinfo
  {author} {\bibfnamefont {Y.-C.}\ \bibnamefont {Zhang}},\ }\href@noop {}
  {\bibfield  {journal} {\bibinfo  {journal} {Physics Reports}\ } (\bibinfo
  {year} {2019})}\BibitemShut {NoStop}%
\bibitem [{\citenamefont {Shin}\ \emph {et~al.}(2016)\citenamefont {Shin},
  \citenamefont {Eliassi-Rad},\ and\ \citenamefont
  {Faloutsos}}]{shin2016corescope}%
  \BibitemOpen
  \bibfield  {author} {\bibinfo {author} {\bibfnamefont {K.}~\bibnamefont
  {Shin}}, \bibinfo {author} {\bibfnamefont {T.}~\bibnamefont {Eliassi-Rad}}, \
  and\ \bibinfo {author} {\bibfnamefont {C.}~\bibnamefont {Faloutsos}},\ }in\
  \href@noop {} {\emph {\bibinfo {booktitle} {2016 IEEE 16th International
  Conference on Data Mining (ICDM)}}}\ (\bibinfo {organization} {IEEE},\
  \bibinfo {year} {2016})\ pp.\ \bibinfo {pages} {469--478}\BibitemShut
  {NoStop}%
\bibitem [{\citenamefont {Balaji}\ \emph {et~al.}(2006)\citenamefont {Balaji},
  \citenamefont {Babu}, \citenamefont {Iyer}, \citenamefont {Luscombe},\ and\
  \citenamefont {Aravind}}]{balaji2006comprehensive}%
  \BibitemOpen
  \bibfield  {author} {\bibinfo {author} {\bibfnamefont {S.}~\bibnamefont
  {Balaji}}, \bibinfo {author} {\bibfnamefont {M.~M.}\ \bibnamefont {Babu}},
  \bibinfo {author} {\bibfnamefont {L.~M.}\ \bibnamefont {Iyer}}, \bibinfo
  {author} {\bibfnamefont {N.~M.}\ \bibnamefont {Luscombe}}, \ and\ \bibinfo
  {author} {\bibfnamefont {L.}~\bibnamefont {Aravind}},\ }\href@noop {}
  {\bibfield  {journal} {\bibinfo  {journal} {Journal of molecular biology}\
  }\textbf {\bibinfo {volume} {360}},\ \bibinfo {pages} {213} (\bibinfo {year}
  {2006})}\BibitemShut {NoStop}%
\bibitem [{\citenamefont {Gama-Castro}\ \emph {et~al.}(2008)\citenamefont
  {Gama-Castro}, \citenamefont {Jim{\'e}nez-Jacinto}, \citenamefont
  {Peralta-Gil}, \citenamefont {Santos-Zavaleta}, \citenamefont
  {Pe{\~n}aloza-Spinola}, \citenamefont {Contreras-Moreira}, \citenamefont
  {Segura-Salazar}, \citenamefont {Muniz-Rascado}, \citenamefont
  {Martinez-Flores}, \citenamefont {Salgado} \emph
  {et~al.}}]{gama2008regulondb}%
  \BibitemOpen
  \bibfield  {author} {\bibinfo {author} {\bibfnamefont {S.}~\bibnamefont
  {Gama-Castro}}, \bibinfo {author} {\bibfnamefont {V.}~\bibnamefont
  {Jim{\'e}nez-Jacinto}}, \bibinfo {author} {\bibfnamefont {M.}~\bibnamefont
  {Peralta-Gil}}, \bibinfo {author} {\bibfnamefont {A.}~\bibnamefont
  {Santos-Zavaleta}}, \bibinfo {author} {\bibfnamefont {M.~I.}\ \bibnamefont
  {Pe{\~n}aloza-Spinola}}, \bibinfo {author} {\bibfnamefont {B.}~\bibnamefont
  {Contreras-Moreira}}, \bibinfo {author} {\bibfnamefont {J.}~\bibnamefont
  {Segura-Salazar}}, \bibinfo {author} {\bibfnamefont {L.}~\bibnamefont
  {Muniz-Rascado}}, \bibinfo {author} {\bibfnamefont {I.}~\bibnamefont
  {Martinez-Flores}}, \bibinfo {author} {\bibfnamefont {H.}~\bibnamefont
  {Salgado}},  \emph {et~al.},\ }\href@noop {} {\bibfield  {journal} {\bibinfo
  {journal} {Nucleic acids research}\ }\textbf {\bibinfo {volume} {36}},\
  \bibinfo {pages} {D120} (\bibinfo {year} {2008})}\BibitemShut {NoStop}%
\bibitem [{\citenamefont {Chung}\ \emph {et~al.}(2003)\citenamefont {Chung},
  \citenamefont {Lu},\ and\ \citenamefont {Vu}}]{chung2003spectra}%
  \BibitemOpen
  \bibfield  {author} {\bibinfo {author} {\bibfnamefont {F.}~\bibnamefont
  {Chung}}, \bibinfo {author} {\bibfnamefont {L.}~\bibnamefont {Lu}}, \ and\
  \bibinfo {author} {\bibfnamefont {V.}~\bibnamefont {Vu}},\ }\href@noop {}
  {\bibfield  {journal} {\bibinfo  {journal} {Proceedings of the National
  Academy of Sciences}\ }\textbf {\bibinfo {volume} {100}},\ \bibinfo {pages}
  {6313} (\bibinfo {year} {2003})}\BibitemShut {NoStop}%
\bibitem [{\citenamefont {Castellano}\ and\ \citenamefont
  {Pastor-Satorras}(2017)}]{castellano2017relating}%
  \BibitemOpen
  \bibfield  {author} {\bibinfo {author} {\bibfnamefont {C.}~\bibnamefont
  {Castellano}}\ and\ \bibinfo {author} {\bibfnamefont {R.}~\bibnamefont
  {Pastor-Satorras}},\ }\href@noop {} {\bibfield  {journal} {\bibinfo
  {journal} {Physical Review X}\ }\textbf {\bibinfo {volume} {7}},\ \bibinfo
  {pages} {041024} (\bibinfo {year} {2017})}\BibitemShut {NoStop}%
\bibitem [{\citenamefont {Sorensen}(1981)}]{sorensen1981interactions}%
  \BibitemOpen
  \bibfield  {author} {\bibinfo {author} {\bibfnamefont {A.}~\bibnamefont
  {Sorensen}},\ }\href@noop {} {\bibfield  {journal} {\bibinfo  {journal}
  {Oecologia}\ }\textbf {\bibinfo {volume} {50}},\ \bibinfo {pages} {242}
  (\bibinfo {year} {1981})}\BibitemShut {NoStop}%
\bibitem [{\citenamefont {Staniczenko}\ \emph {et~al.}(2013)\citenamefont
  {Staniczenko}, \citenamefont {Kopp},\ and\ \citenamefont
  {Allesina}}]{staniczenko2013ghost}%
  \BibitemOpen
  \bibfield  {author} {\bibinfo {author} {\bibfnamefont {P.~P.}\ \bibnamefont
  {Staniczenko}}, \bibinfo {author} {\bibfnamefont {J.~C.}\ \bibnamefont
  {Kopp}}, \ and\ \bibinfo {author} {\bibfnamefont {S.}~\bibnamefont
  {Allesina}},\ }\href@noop {} {\bibfield  {journal} {\bibinfo  {journal}
  {Nature communications}\ }\textbf {\bibinfo {volume} {4}},\ \bibinfo {pages}
  {1} (\bibinfo {year} {2013})}\BibitemShut {NoStop}%
\bibitem [{\citenamefont {Mariani}\ \emph {et~al.}(2019)\citenamefont
  {Mariani}, \citenamefont {Ren}, \citenamefont {Bascompte},\ and\
  \citenamefont {Tessone}}]{mariani2019nestedness}%
  \BibitemOpen
  \bibfield  {author} {\bibinfo {author} {\bibfnamefont {M.~S.}\ \bibnamefont
  {Mariani}}, \bibinfo {author} {\bibfnamefont {Z.-M.}\ \bibnamefont {Ren}},
  \bibinfo {author} {\bibfnamefont {J.}~\bibnamefont {Bascompte}}, \ and\
  \bibinfo {author} {\bibfnamefont {C.~J.}\ \bibnamefont {Tessone}},\
  }\href@noop {} {\bibfield  {journal} {\bibinfo  {journal} {Physics Reports}\
  } (\bibinfo {year} {2019})}\BibitemShut {NoStop}%
\bibitem [{\citenamefont {Lever}\ \emph {et~al.}(2014)\citenamefont {Lever},
  \citenamefont {van Nes}, \citenamefont {Scheffer},\ and\ \citenamefont
  {Bascompte}}]{lever2014sudden}%
  \BibitemOpen
  \bibfield  {author} {\bibinfo {author} {\bibfnamefont {J.~J.}\ \bibnamefont
  {Lever}}, \bibinfo {author} {\bibfnamefont {E.~H.}\ \bibnamefont {van Nes}},
  \bibinfo {author} {\bibfnamefont {M.}~\bibnamefont {Scheffer}}, \ and\
  \bibinfo {author} {\bibfnamefont {J.}~\bibnamefont {Bascompte}},\ }\href@noop
  {} {\bibfield  {journal} {\bibinfo  {journal} {Ecology letters}\ }\textbf
  {\bibinfo {volume} {17}},\ \bibinfo {pages} {350} (\bibinfo {year}
  {2014})}\BibitemShut {NoStop}%
\bibitem [{\citenamefont {Rohr}\ \emph {et~al.}(2014)\citenamefont {Rohr},
  \citenamefont {Saavedra},\ and\ \citenamefont
  {Bascompte}}]{rohr2014structural}%
  \BibitemOpen
  \bibfield  {author} {\bibinfo {author} {\bibfnamefont {R.~P.}\ \bibnamefont
  {Rohr}}, \bibinfo {author} {\bibfnamefont {S.}~\bibnamefont {Saavedra}}, \
  and\ \bibinfo {author} {\bibfnamefont {J.}~\bibnamefont {Bascompte}},\
  }\href@noop {} {\bibfield  {journal} {\bibinfo  {journal} {Science}\ }\textbf
  {\bibinfo {volume} {345}},\ \bibinfo {pages} {1253497} (\bibinfo {year}
  {2014})}\BibitemShut {NoStop}%
\bibitem [{\citenamefont {Saavedra}\ \emph {et~al.}(2016)\citenamefont
  {Saavedra}, \citenamefont {Rohr}, \citenamefont {Olesen},\ and\ \citenamefont
  {Bascompte}}]{saavedra2016nested}%
  \BibitemOpen
  \bibfield  {author} {\bibinfo {author} {\bibfnamefont {S.}~\bibnamefont
  {Saavedra}}, \bibinfo {author} {\bibfnamefont {R.~P.}\ \bibnamefont {Rohr}},
  \bibinfo {author} {\bibfnamefont {J.~M.}\ \bibnamefont {Olesen}}, \ and\
  \bibinfo {author} {\bibfnamefont {J.}~\bibnamefont {Bascompte}},\ }\href@noop
  {} {\bibfield  {journal} {\bibinfo  {journal} {Ecology and evolution}\
  }\textbf {\bibinfo {volume} {6}},\ \bibinfo {pages} {997} (\bibinfo {year}
  {2016})}\BibitemShut {NoStop}%
\bibitem [{\citenamefont {Pastor-Satorras}\ \emph {et~al.}(2015)\citenamefont
  {Pastor-Satorras}, \citenamefont {Castellano}, \citenamefont {Van~Mieghem},\
  and\ \citenamefont {Vespignani}}]{pastor2015epidemic}%
  \BibitemOpen
  \bibfield  {author} {\bibinfo {author} {\bibfnamefont {R.}~\bibnamefont
  {Pastor-Satorras}}, \bibinfo {author} {\bibfnamefont {C.}~\bibnamefont
  {Castellano}}, \bibinfo {author} {\bibfnamefont {P.}~\bibnamefont
  {Van~Mieghem}}, \ and\ \bibinfo {author} {\bibfnamefont {A.}~\bibnamefont
  {Vespignani}},\ }\href@noop {} {\bibfield  {journal} {\bibinfo  {journal}
  {Reviews of modern physics}\ }\textbf {\bibinfo {volume} {87}},\ \bibinfo
  {pages} {925} (\bibinfo {year} {2015})}\BibitemShut {NoStop}%
\bibitem [{\citenamefont {Berman}\ and\ \citenamefont
  {Plemmons}(1994)}]{berman1994nonnegative}%
  \BibitemOpen
  \bibfield  {author} {\bibinfo {author} {\bibfnamefont {A.}~\bibnamefont
  {Berman}}\ and\ \bibinfo {author} {\bibfnamefont {R.~J.}\ \bibnamefont
  {Plemmons}},\ }\href@noop {} {\emph {\bibinfo {title} {Nonnegative matrices
  in the mathematical sciences}}}\ (\bibinfo  {publisher} {SIAM},\ \bibinfo
  {year} {1994})\ p.~\bibinfo {pages} {27}\BibitemShut {NoStop}%
\bibitem [{\citenamefont {Collatz}(1942)}]{collatz1942einschliessungssatz}%
  \BibitemOpen
  \bibfield  {author} {\bibinfo {author} {\bibfnamefont {L.}~\bibnamefont
  {Collatz}},\ }\href@noop {} {\bibfield  {journal} {\bibinfo  {journal}
  {Mathematische Zeitschrift}\ }\textbf {\bibinfo {volume} {48}},\ \bibinfo
  {pages} {221} (\bibinfo {year} {1942})}\BibitemShut {NoStop}%
\bibitem [{\citenamefont {Wielandt}(1950)}]{wielandt1950unzerlegbare}%
  \BibitemOpen
  \bibfield  {author} {\bibinfo {author} {\bibfnamefont {H.}~\bibnamefont
  {Wielandt}},\ }\href@noop {} {\bibfield  {journal} {\bibinfo  {journal}
  {Mathematische Zeitschrift}\ }\textbf {\bibinfo {volume} {52}},\ \bibinfo
  {pages} {642} (\bibinfo {year} {1950})}\BibitemShut {NoStop}%
\end{thebibliography}%
\section*{Methods}

\subsection{Properties}

We study the systems that follow the dynamic equation:

\begin{equation}
\frac{dx_i}{dt}=F(x_i, \sum_{j=1}^{N}A_{ij}G(x_i,x_j)),
\label{dynamical}
\end{equation}
$x_i \geq 0$ is the abundance of components $i$, weighted $A_{ij} \geq 0$ measures the interaction strength from $j$ to $i$. The cooperative systems we study in this paper have the following properties:

(i) $F(\mathbf{x=0}) = 0$. Assume $F(\mathbf{x=0}) > 0$, then obviously there exist a nonzero solution. And abundance cannot be negative, so it is impossible that $F(\mathbf{x=0}) < 0$.

(ii) Abundance can only be zero if there is no input: $F(x_i>0,0)<0$; 

(iii) More input lead to higher fixed-point abundance: $\partial F/\partial u_i \geq 0$, here $u_i = \sum_{j=1}^{N}A_{ij}G(x_i,x_j)$ is the second variable of $F$;

(iv) Higher abundance produce more output: $\frac{\partial G}{\partial x_j} \geq 0$, and obviously, $G(x_i,0) = 0$;
 
(v) $\mathbf{x}$ cannot increase to infinite, $\lim\limits_{\Vert x\Vert\to+\infty} \frac{d \Vert x\Vert}{dt} < 0$.

Among them, (i), (ii), and (v) are generally valid for most of the dynamic systems. 

\subsection{Conditions}
Here we study the following three conditions: 

(i) The system $\mathcal{A}$ follows Eq.~\ref{dynamical} have nonzero solution;

(ii) A homogeneous system with total weighted in-degree equals to $k_{max}(A)$ has positive solution. In another word, 1D equation $F(x, k_{max}(A) G(x,x)) = 0$ has positive solution $x_k^*$;

(iii)  A homogeneous system with total weighted in-degree equals to $\rho(A)$ has positive solution. In another word, 1D equation $F(x, \rho(A) G(x,x)) = 0$ has positive solution $x_\rho^*$.

We will show that, condition (ii) is the sufficient condition of condition (i), condition (iii) is the necessary condition of condition (i); when $G(x_i,x_j)$ is a step function of $x_j$, condition (ii) is equivalent to condition(i); and when $F$ from zero solution to nonzero solution undergoes a continuous transition, condition (iii) is equivalent to condition (i). 

\subsection{Lemmas}
Here we introduce two important lemmas we will use later:

(i)
$A \geq 0$, $B \geq 0$, and $B_{ij} \leq A_{ij}$, $\forall i, j$, then
$\rho(B) \leq \rho(A)$.~\cite{berman1994nonnegative}

(ii) Collatz–Wielandt theorem~\cite{collatz1942einschliessungssatz, wielandt1950unzerlegbare}:
$A \geq 0 $, $\forall \mathbf{x} > \mathbf{0}$, then
\begin{equation}
\min_{1\leq i \leq n} \frac{A_{ij}x_i}{x_i} \leq \rho(A) \leq \max_{1\leq i \leq n} \frac{A_{ij}x_i}{x_i}
\end{equation}

\subsection{Reasoning from condition (ii) to condition (i)}
By definition of k-core, we can delete the nodes and decrease the weight of edges in system $\mathcal{A}$ to construct a system $\mathcal{B}$ that every node has total weighted in-degree equals to $k_{max}(A)$, we denote the nodes in $\mathcal{B}$ by $1,2,...,n$. Obviously, $x_1 = x_2 =...= x_k^*$ is a solution of system $\mathcal{B}$. Obviously the corresponding nodes $1, 2,..., n$ in system $\mathcal{A}$ must have a solution no less than $x_k^*$.

\subsection{Reasoning from condition (i) to condition (iii)}
The system $\mathcal{A}$ has nonzero solution 
indicates that there exist some effective interactions $G(x_i^*,x_j^*)>0$,
we remove all the nodes that have zero abundance, and all the edges that have zero influence $G(x_i^*,x_j^*) = 0$.
We denote the remaining nodes as $1,2,...,n$, and the ''effective'' adjacency matrix $B_{n*n}$.
Obviously, in system $\mathcal{B}$, $x_i^*$ will stay the same as it in system $\mathcal{A}$, and $x_i^* > 0$, $\forall 1 \leq i \leq n$. We have 
\begin{equation}
\sum_{j=1}^{N}A_{ij}G(x_i^*,x_j^*) = \sum_{j=1}^{n}B_{ij}G(x_i^*,x_j^*)
\end{equation}
By Lemma (i), (ii) and 
$\rho(B) = \rho(
\left[
 \begin{matrix}
   B & 0\\
   0 & 0
  \end{matrix}
  \right] 
  )
$
, there exists component $i$ ($1 \leq i \leq n$) such that
\begin{equation}
\sum_{j=1}^{n}B_{ij}G(x_i^*,x_j^*)/G(x_i^*,x_i^*) \leq \rho(B) \leq \rho(A).
\end{equation}
Hence
\begin{align}
\frac{dx_i}{dt}\bigg{\arrowvert}_{x_i^*}&=F(x_i^*, \sum_{j=1}^{N}A_{ij}G(x_i^*,x_j^*))\\
&=F(x_i^*, \sum_{j=1}^{n}B_{ij}G(x_i^*,x_j^*))\\
&\leq F(x_i^*, \rho(A) G(x_i^*,x_i^*))
\end{align}
$dx_i^*/dt = 0$, then
\begin{equation}
F(x_i^*, \rho(A) G(x_i^*,x_i^*)) \geq 0.
\end{equation}
$F(x, \rho(A) G(x,x)) = 0$ must have solution no less than $x_i^*$.

\subsection{Reasoning from condition (i) to condition (ii)}
When $G(x_i,x_j)$ is a step function:
\begin{equation}
G(x_i,x_j) =
\left	\{
             \begin{array}{ll}
             0 & \textrm{if  $x_j < \alpha$}   \\
             g(x_i)  & \textrm{if  $x_j \geq \alpha$}
             \end{array}
\right.
\end{equation}
The system has nonzero solutions indicates the existence of a group of ''symbionts'' whose abundance are all larger than $\alpha$. 
We denote their id as $1,2,...,n$ and the interaction matrix within the group is $B$.
The fixed-point solution
\begin{equation}
F(x_i^*, \sum_{j=1}^{N}A_{ij}G(x_i^*,x_j^*)) = F(x_i^*, \sum_{j=1}^{n}B_{ij} g(x_i^*)) = 0
\end{equation}
$\sum B_{mj}$ is the smallest among all $\sum B_{ij}$, $1\leq i \leq n$
\begin{equation}
F(x_m^*, \sum_{j=1}^{n}B_{mj} g(x_m^*)) = 0.
\end{equation}
By definition, $\sum_{j=1}^{n}B_{mj}  \leq k_{max}(A)$, therefore
\begin{equation}
F(x_m^*,k_{max}(A)g(x_m^*)) \geq 0
\end{equation}
Therefore, $F(x,k_{max}(A)g(x)) = 0$ must have a solution $x^*$ no less than $x_m^*$.
On the other hand $x_m^* \geq \alpha$, hence $F(x^*,k_{max}(A)g(x^*))$ is equivalent to $F(x^*, k_{max}(A) G(x^*,x^*))$. 
 
\subsection{Reasoning from condition (iii) to condition (i)}
If the graph is not strongly connected, that is to say we can divide it to different strongly connected subgraphs, and there is no feedback loop among different subgraphs. Obviously, the system will have a nonzero solution if any of these strongly connected subgraphs in isolation has a nonzero solution.

On the other hand, we can always write the adjacency matrix in a block triangular form, and easily prove that the largest eigenvalue of the whole graph equals to the largest one among all the largest eigenvalues of these strongly connected subgraphs. 

The adjacency matrix $A$ of a strongly connected subgraph is an irreducible matrix, Perron–Frobenius Theory guarantees that $\rho(A)$ is positive, and the dominant eigenvector such that
$\mathbf{\omega}^TA=\rho(A)\mathbf{\omega}^T$
is positive. We will use this property in the following proof.

In what follows we use Lyapunov’s second method to illustrate the local stability at $\mathbf{x} = \mathbf{0}$. Note $A$ is a irreducible matrix in the following proof.

\subsubsection{Gene regulatory network}
\begin{equation}
\frac{dx_i}{dt}=-x_i^f+\sum_{j=1}^{N}\widetilde{A}_{ij}\frac{x_j^h}{1+x_j^h},
\label{ge_eq}
\end{equation}
near $\mathbf{x} = \mathbf{0}$,
\begin{equation}
\frac{dx_i}{dt}=-x_i^f+\sum_{j=1}^{N}\widetilde{A}_{ij} x_j^h.
\end{equation}
The corresponding 1D equation
\begin{equation}
\frac{dx}{dt}=-x^f+\rho(\widetilde{A}) x^h.
\end{equation}
Obviously, when $f<h$, it is locally stable at $x = 0$, and its nonzero solution cannot be studied near $x = 0$; when $f>h$, it is unstable at $x = 0$; the two results are trivial.
When $f = h$, the 1D equation is unstable at $x = 0$ if and only if $\rho(\widetilde{A}) > 1$. 

Then we study the complex system $\mathcal{A}$ when $f = h$. Choose Lyapunov function $V(\mathbf{x}) = \mathbf{\omega} \cdot \mathbf{x}$, here $\mathbf{\omega}$ is the dominant eigenvector of $A$ such that $\mathbf{\omega}^TA=\rho(A)\mathbf{\omega}^T$.  Obviously $V(\mathbf{x})>0$ when $\mathbf{x} \neq 0$, and
\begin{equation}
\frac{dV(\mathbf{x})}{dt} = \mathbf{\omega} \cdot \frac{d\mathbf{x}}{dt}
= (\rho(\widetilde{A})-1) \sum_{j=1}^{N}w_j x_j^h.
\end{equation}
When $\rho(\widetilde{A})>1$, $\frac{dV(\mathbf{x})}{dt}>0$ when $\mathbf{x}$ near $\mathbf{0}$ but $\mathbf{x} \neq \mathbf{0}$. The system is unstable at $\mathbf{x} = \mathbf{0}$, thus the system follows Eq.~\ref{ge_eq} must have a nonzero solution.
Condition (iii) can lead to condition (i) when $f \geq h$. Especially, when $f = h$, we have $\lambda_c^* = \rho(A)$.

\subsubsection{Mutualistic network}
Then we study the mutualistic network ruled by the following dynamic equation:
\begin{equation}
\frac{dx_i}{dt}=-x_id-x_i^2s+\frac{ \gamma \sum_{j=1}^{N} A_{ij} x_j }{ \alpha + \sum_{j=1}^{N}A_{ij}x_j}x_i,
\end{equation}
when $d = 0$,
\begin{equation}
\frac{dx_i}{dt}=(-x_i s+\frac{ \gamma \sum_{j=1}^{N} A_{ij} x_j }{ \alpha + \sum_{j=1}^{N}A_{ij}x_j})x_i,
\end{equation}
its nonzero solution is equivalent to the solution of the following system
\begin{equation}
\frac{dx_i}{dt}=-x_i s+\frac{ \gamma \sum_{j=1}^{N} A_{ij} x_j }{ \alpha + \sum_{j=1}^{N}A_{ij}x_j}.
\end{equation}
Near $\mathbf{x} = \mathbf{0}$,
\begin{equation}
\frac{dx_i}{dt}=-x_i s+\frac{\gamma}{\alpha} \sum_{j=1}^{N} A_{ij} x_j,
\end{equation}
The rest part of proof is similar to the case of gene regulatory networks.
We have the conclusion that when $d = 0$,  $\lambda_c^* = \rho(A)$.

\subsubsection{SIS model}
\begin{equation}
\frac{dx_i}{dt} = -x_i  + (1-x_i) \sum_{j=1}^{N} \widetilde{A}_{ij}x_j,
\end{equation}
near $\mathbf{x} = \mathbf{0}$,
\begin{equation}
\frac{dx_i}{dt} = -x_i  + \sum_{j=1}^{N} \widetilde{A}_{ij}x_j.
\end{equation}
Again, similarly to the case of gene regulatory networks, we have the conclusion that the SIS model always has $\lambda_c^* = \rho(A)$.

\subsubsection{general}
In general, we conjecture that when the transition from zero solution to positive solution of $1D$ equation 
\begin{equation}
F(x, \lambda G(x,x)) = 0
\end{equation}
 is continuous, the original system follows Eq.~\ref{dynamical} has nonzero solution is equivalent to that the homogeneous system with total weighted in-degree equals to $\rho(A)$ has a positive solution.

\end{document}